\documentclass{article}

\usepackage[dblblindworkshop,nonatbib,final]{neurips_2026}

\usepackage[utf8]{inputenc}
\usepackage[T1]{fontenc}

\usepackage[pdfpagelabels=false,colorlinks=true,allcolors=blue]{hyperref}
\usepackage{url}
\usepackage{booktabs}
\usepackage{amsfonts}
\usepackage{amssymb}
\usepackage{nicefrac}
\usepackage{microtype}
\usepackage{xcolor}         
\usepackage{amsmath}
\usepackage{graphicx}
\usepackage{comment}
\usepackage{multirow}
\usepackage{tabularx}
\usepackage{aas_macros}     
\usepackage[backend=bibtex,style=numeric-comp,natbib=true,sorting=none,
            maxbibnames=1,minbibnames=1,maxcitenames=2,giveninits=true,
            url=false,doi=false,isbn=false,eprint=false]{biblatex}
\newcommand{\arrup}{\rlap{$\,\uparrow$}}
\newcommand{\arrdown}{\rlap{$\,\downarrow$}}

\graphicspath{{./}{figures/}}

\workshoptitle{Representations for the Physical Sciences Workshop}

\title{Self-Supervised Representation Learning for JWST Galaxy Images and Low Resolution Spectra}

  \author{
  Yang Cheng \\
  Max-Planck-Institut f\"ur Astronomie (MPIA) \\
  K\"onigstuhl 17, 69117 Heidelberg, Germany \\
  Heidelberg University, Heidelberg, Germany \\
  \texttt{yacheng@mpia.de} \\
  \And
  Morgan Fouesneau \\
  Max-Planck-Institut f\"ur Astronomie (MPIA) \\
  K\"onigstuhl 17, 69117 Heidelberg, Germany \\
  \texttt{fouesneau@mpia.de} \\
  \And
  Ivelina G. Momcheva \\
  Max-Planck-Institut f\"ur Astronomie (MPIA) \\
  K\"onigstuhl 17, 69117 Heidelberg, Germany \\
  \texttt{momcheva@mpia.de} \\
  \And
  Tobias Buck \\
  Interdisciplinary Center for Scientific Computing (IWR) \\
  Heidelberg University \\
  Im Neuenheimer Feld 205, 69120 Heidelberg, Germany \\
  \texttt{tobias.buck@iwr.uni-heidelberg.de} \\
  \And
  Shengxiu Sun \\
  Department of Astronomy, School of ßPhysics, Peking University \\
  Beijing 100871, China \\
  Kavli Institute for Astronomy and Astrophysics, Peking University \\
  Beijing 100871, China \\
  Institute for Theoretical Physics, Heidelberg University \\
  Philosophenweg 12, 69120 Heidelberg, Germany \\
  \texttt{sxsun@stu.pku.edu.cn} \\
}

\begin{document}

\maketitle

\begin{abstract}

A self-supervised multimodal model recovers galaxy stellar masses from sparse \emph{JWST} data that lack the multi-band coverage required by traditional SED fitting. We train on single-band F150W imaging paired with observed-frame $1$--$2~\mu$m PRISM spectra and obtain a joint image-spectrum representation that recovers stellar mass at $1<z\le3$ with $R^2 \sim 0.9$ and $\sigma_{\mathrm{NMAD}} = 0.24$~dex; cross-modal CLIP alignment further raises single-modality performance by $\Delta R^2 = +0.04$ (imaging) and $+0.09$ (spectroscopy), narrowing the gap to joint-modality inference. The code and documentation are available at 
\href{https://github.com/2003cy/neurips2026_ssl_jwst}{GitHub}.
\end{abstract}


\section{Introduction} \label{sec:intro}

Extragalactic parameter inference relies on translating observations into physical properties. Spectral energy distribution (SED) fitting \citep[e.g.,][]{intro1, intro2, intro3, intro4, intro5} achieves this through broad-band wavelength coverage that breaks degeneracies between stellar age, dust, and metallicity \citep{eazy, eazypy, prospector}. Space-based surveys such as \emph{JWST} and the upcoming \emph{Roman Space Telescope} now reach wider fields, but individual sources increasingly lack the multi-band observations needed to break these degeneracies \citep{sorba2018, pforr2012, zimmerman1, jones2026}.

Self-supervised multimodal models learn representations from unlabeled data that capture empirical structures in the observed population, rather than relying on human-imposed priors. Recent work matches traditional SED fitting from multi-band imaging \citep{astroclip2023, astropt, aion2025}, but uses complete ground-based datasets such as the DESI Legacy Imaging Surveys. Space-based observations with sparse coverage, despite their growing relevance, remain unexplored in this setting.

This study trains a framework on a restricted \emph{JWST} dataset of single-band F150W imaging and observed-frame $1$--$2~\mu$m \emph{NIRSpec} low-resolution spectra. Two single-modality encoders are pre-trained and then aligned by a cross-modal CLIP head; linear probes evaluate the resulting representations for downstream parameters (redshift, stellar mass) against multi-band SED fitting and recover stellar mass at $R^2 \approx 0.90$ ($\sigma_{\mathrm{NMAD}} = 0.24$~dex) at $1<z\le3$ (Section~\ref{sec:results}).

\section{Data} \label{sec:data}

We construct a restricted dataset from multiple \emph{JWST} archives to emulate the sparse band coverage of wide-field surveys. The technical details and data preparation are described in Appendix~\ref{app:data}.

\textbf{Imaging (\ref{app:image}).} We use single-band \emph{JWST} F150W observations from the OutThere pure-parallel survey (NIRISS) and the NIRCam surveys COSMOS-Web \citep{cosmos}, JADES \citep{jades}, and CEERS \citep{ceers}. Cutouts are extracted at $128\times128$ pixels ($30$ milliarcseconds per pixel, $\mathrm{MJy\,sr^{-1}}$ units) and center-cropped to $100\times100$ after augmentation. We pre-select based on each survey's own quality or bad-region flags and a positive integrated signal-to-noise ratio so that faint or contaminated sources are removed. We also discard cutouts with more than $40\%$ invalid pixels from detector edges or artifacts. In total, we retain $\sim 1.4 \times 10^{6}$ cutouts.

\textbf{Spectra (\ref{app:spectra}).} We use spectral data from the DAWN \emph{JWST} Archive (DJA) \citep{Heintz2024, degraff2025}, which provides uniformly processed \emph{JWST}/NIRSpec observations reduced with the \texttt{msaexp} pipeline \citep{msaexp}. We use the low-resolution PRISM/CLEAR spectra and
retain only the observed-frame $1$--$2~\mu$m window ($R\sim100$, a fixed grid of $56$ pixels), emulating both a sparse spectral setup and the wavelength coverage of the \emph{Roman} grism ($1.00$--$1.93~\mu$m) \citep{akeson2019} and the \emph{Euclid} NISP red grisms ($1.206$--$1.892~\mu$m) \citep{euclid2025}.
We further filter by the DJA quality grade and require at least $50\%$ of the pixels in this wavelength window to be valid. This leaves a total of $25{,}025$ spectra. The matched spectra span redshifts up to $z\sim7$.

\textbf{Cross-matching (\ref{app:match}).} We pair spectra to imaging by sky position within $0.3''$, using imaging as the reference because both modalities contain repeat observations of the same source through overlapping NIRCam/NIRISS pointings or multiple PRISM observations. The $11{,}664$ matched image-spectrum pairs ($10{,}925$ unique spectra) are grouped by unique DJA source identifier so that multiple cutouts of one source never leak across train/test splits during alignment and downstream tasks.


\section{Model} \label{sec:method}

\begin{figure}[h]
    \centering
    \includegraphics[width=1\linewidth]{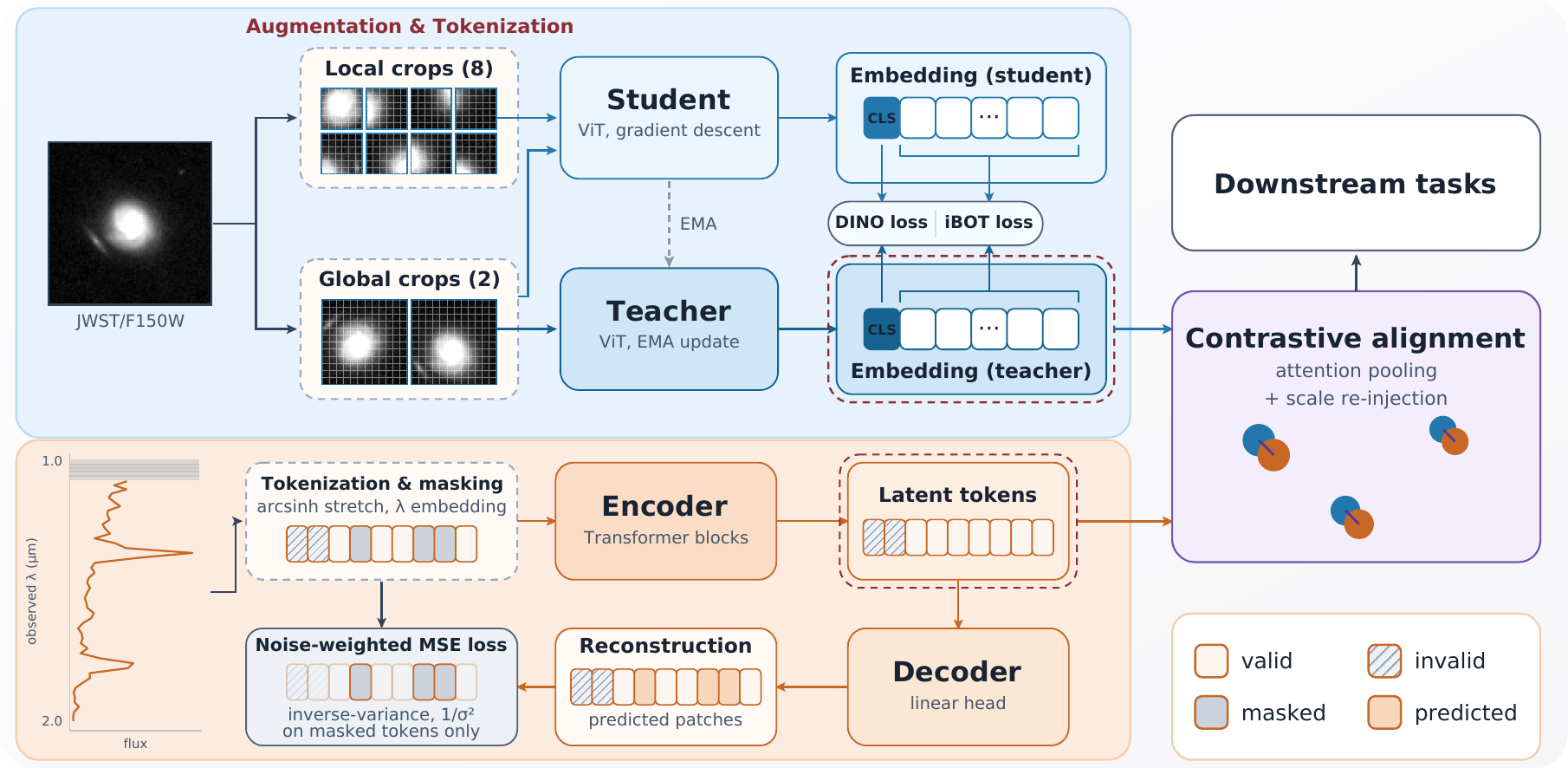}
    \caption{Top left: a ViT trained with DINOv2 by aligning teacher and student outputs from augmented views of one image. Bottom left: the spectrum encoder, a 1D autoencoder, tokenizes the input spectrum into overlapping flux patches (masked tokens shown as gray squares) and reconstructs them via a noise-weighted loss. Invalid tokens (hatched squares) are excluded end to end. Right: CLIP alignment after pre-training. The image and spectrum model outputs are attention-pooled, the scale is re-injected, and they are aligned with a contrastive objective.}
    \label{fig:model}
\end{figure}

Our framework has two modality-specific encoders pre-trained independently in a self-supervised manner on the full imaging and spectroscopic samples (Figure~\ref{fig:model}). A CLIP-style head then aligns their frozen outputs into a shared latent space using the cross-matched pairs.

\textbf{Image encoder (\ref{app:model_img}).} We pre-train a single-channel vision transformer \citep[ViT;][]{dosovitskiy2021} via DINOv2 \citep{oquab2024} (upper-left panel of Figure~\ref{fig:model}), which has been applied in astronomy \citep{astroclip2023, stradavit, Lastufka2025}. Each image yields several global and local augmented views, tokenized into overlapping patch tokens with a learnable \texttt{[CLS]} token as the global representation. A student network receives all views, while the teacher with the same architecture sees the global views. The training combines two objectives: the DINO loss, regularized by KoLeo, aligns the \texttt{[CLS]} token between the two networks, and the iBOT loss \citep{ibot} predicts the student's tokens under random patch masking. The student is optimized with the AdamW optimizer \citep{adamw}; the teacher parameters follow an exponential moving average (EMA) of the student. The augmentation pipeline and the  details of the architecture are described in Appendix~\ref{app:model_img}.

\textbf{Spectrum encoder (\ref{app:model_spec}).} We train a transformer autoencoder with a masked-modeling objective \citep{he_masked_2021}, following
OmniSpectra \citep{omnispectra2026}. An architecture summary is shown in the bottom left panel of Figure~\ref{fig:model}. Each spectrum is $\mathrm{arcsinh}$-stretched and standardized, then tokenized into overlapping flux patches (Appendix~\ref{app:model_spec}); tokens with any invalid pixels are excluded end to end, and the rest are randomly masked in chunks. The sequence is passed to the transformer encoder and a single linear decoder. Since the pixel signal-to-noise ratio varies strongly along a spectrum, we weight each token $t$ by $w_t = 1/\max(\sigma_t^2, \sigma^2_{\min})$ at masked positions, with $\sigma_t^2$ propagated from the per-pixel noise and a parameter $\sigma^2_{\min}$ preventing a few clean tokens from dominating. The final loss is a noise-weighted mean squared error (MSE) over the masked tokens $M$, $\mathcal{L}_{\text{MSE}} = \sum_{t\in M}\tilde{w}_t\left(\sum_{i\in t}(\hat{x}_i-x_i)^2\right)$, where $x_i$ and $\hat{x}_i$ are the observed and reconstructed standardized flux of pixel $i$ and $\tilde{w}_t = w_t/\sum_{t\in M} w_t$.

\textbf{CLIP alignment (\ref{app:model_clip}).} Following previous works \citep{buck,astroclip2023,clip}, we map both modality embeddings into a shared 256-dimensional latent space with CLIP. Each token sequence is collapsed into a fixed-length vector by attention pooling \citep{lee2019}, where a learnable query returns a single pooled vector. We re-inject the per-modality scale by adding the arcsinh-stretched mean and standard deviation to the pooled vector in the CLIP head, and concatenate the same statistics to the unaligned embeddings in the downstream tasks so that the probes read identical information before and after alignment. The pooled embeddings are aligned with an InfoNCE objective \citep{oord2018, clip}, pulling matched image-spectrum pairs together and pushing all other pairs apart.

\begin{figure}[ht]
    \centering
    \includegraphics[width=1\linewidth]{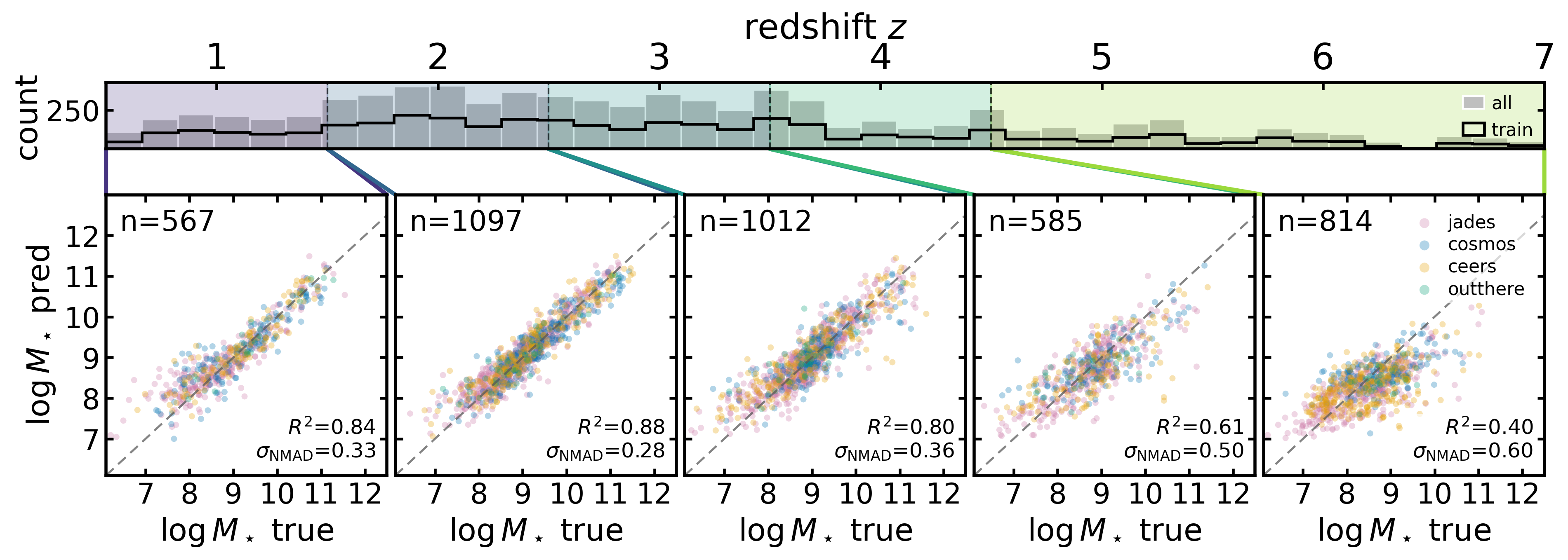}
    \caption{Stellar-mass linear probe on the CLIP-aligned image-spectrum embedding. Top: redshift distribution up to $z\sim7$ of the labeled pairs (gray) with the training half overlaid (black step) and the five evaluation bins shaded. Bottom: predicted versus catalog masses of the test set, colored by imaging survey; the probe is trained once and only its predictions are binned. Dashed line: one-to-one relation. The points are the test half of a single split (seed 42), whereas the quoted $R^2$ and $\sigma_{\mathrm{NMAD}}$ are the mean over the $40$ splits; their split-to-split standard deviations are at most $0.011$ and $0.009$~dex, and $0.7\%$ for the outlier fraction at $1<z\le3$.}
    \label{fig:mass}
\end{figure}

\section{Results} \label{sec:results}

\textbf{Linear probe.} We benchmark each embedding setup by regressing physical parameters using a single linear transformation on the frozen embeddings, following \citet{astroclip2023, omnispectra2026}. We adopt a ridge ($L_2$) linear regression to map frozen representation to the physical labels, using $40$ random $50\%/50\%$ train/test splits. We report the mean coefficient of determination $R^2$, normalized median absolute deviation $\sigma_{\mathrm{NMAD}}$, and the outlier fraction with $|\Delta\log M_\star| > 0.5$~dex, with split-to-split uncertainties (Appendix~\ref{app:bench}).

\textbf{Single encoder representations.} The pre-trained single encoders recover physically meaningful structure: the linear probe gives a mean $R^2 = 0.75$ (image encoder) across the three S\'ersic parameters ($n$, $R_e$, $q$) and $R^2 = 0.87$ (spectrum encoder) for catalog redshift over $1<z\le3$. Full benchmarks are in Appendix~\ref{app:bench}; the encoders thus capture physically meaningful properties of both modalities without fine-tuning with labeled data.

\textbf{Stellar mass probe.} Stellar mass labels come from multi-band \texttt{EAZY} SED fitting in the DJA catalog; we keep the $7{,}621$ of the $11{,}664$ matched image-spectrum pairs with a valid mass, a valid redshift, and $|z_{\mathrm{phot}}-z_{\mathrm{spec}}|/(1+z_{\mathrm{spec}}) < 0.15$. We train one probe on the filtered set and evaluate its test predictions in binned redshift, shown in Figure~\ref{fig:mass}. Performance depends strongly on redshift: stable below $z\simeq3.5$, strongest at $1.5<z<2.5$ ($R^2=0.87$, with $\sigma_{\mathrm{NMAD}}=0.29$~dex), then it drops at $z>3.5$ to $R^2=0.61$ and $0.40$ in the last two bins. This is notable because the 56-pixel PRISM spectrum has a narrower baseline than a typical photometric set. At $z>3.5$, this degradation occurs because the F150W band and the PRISM window sample the rest-frame ultraviolet, which correlates more with recent star formation rather than with assembled stellar mass.

\textbf{Mass estimation at $1<z\le3$.} Table~\ref{tab:mass} compares stellar-mass probes across data configurations in this range, using the same strategy before and after CLIP alignment 
against two linear baselines: F150W photometry alone, and F150W photometry combined with redshift.
All probes train and test on the $3{,}452$ pairs filtered to $1< z \le 3$. The joint image-spectrum representation reaches $R^2 = 0.90$ in stellar mass, a gain of $0.15$ over the photometric baseline alone, and
$\sim0.1$ over F150W photometry with redshift. Every learned representation improves on the photometric baseline in $R^2$. CLIP alignment raises single-modality performance by $\Delta R^2 = +0.038 \pm 0.006$ for imaging and $+0.092 \pm 0.007$ for spectroscopy, and lowers the outlier fraction from $20.1\%$ to $17.1\%$ (imaging) and from $22.4\%$ to $13.4\%$ (spectroscopy). 
The joint embedding exceeds the photometry + redshift baseline in all metrics, giving the best performance overall ($R^2 \simeq 0.90$, $\sigma_{\mathrm{NMAD}} \simeq 0.24$~dex). 
Yet for the joint embedding, the alignment itself makes no difference within the split-to-split scatter. We conjecture this discrepancy reflects an information ceiling: the concatenated representation already decodes stellar mass equally well before and after alignment. Appendix~\ref{app:breakdown} fits the aligned and unaligned joint embedding on redshift, F150W photometry, and mass.

\begin{table}[t]
\centering
\caption{Stellar-mass linear probes at $1<z\le3$ ($3{,}452$ sources, $40$ splits; ``un.''/``al.'' = "unaligned/aligned", i.e., before/after CLIP alignment). Metrics: $R^2$, $\sigma_{\mathrm{NMAD}}$ [dex], and the $|\Delta\log M_\star|>0.5$~dex fraction; split-to-split scatters are at most $0.011$, $0.009$~dex, and $0.7\%$ respectively.}
\label{tab:mass}

\setlength{\tabcolsep}{9pt}
\begin{tabular}{lcccccc}
\toprule
Representation & \multicolumn{2}{c}{$R^2$} & \multicolumn{2}{c}{$\sigma_{\mathrm{NMAD}}$ [dex]}
& \multicolumn{2}{c}{$f_{>0.5}$ [\%]} \\
\midrule
F150W photometry & \multicolumn{2}{c}{$0.750$} & \multicolumn{2}{c}{$0.358$}
& \multicolumn{2}{c}{$19.5$} \\
F150W photometry + $z$ & \multicolumn{2}{c}{$0.807$} & \multicolumn{2}{c}{$0.275$}
& \multicolumn{2}{c}{$13.8$} \\
\midrule
 & \textit{un.} & \textit{al.} & \textit{un.} & \textit{al.} & \textit{un.} & \textit{al.} \\
\midrule
Image          & $0.780$ & $0.818$\arrup & $0.376$ & $0.342$\arrdown & $20.1$ & $17.1$\arrdown \\
Spectrum         & $0.764$ & $0.856$\arrup & $0.391$ & $0.303$\arrdown & $22.4$ & $13.4$\arrdown \\
Image + spectrum & $0.896$ & $0.902$     & $0.242$ & $0.244$     & $8.1$  & $8.7$ \\
\bottomrule
\end{tabular}
\end{table}

\textbf{Literature comparison.} Prior SED-fitting studies at similar wavelengths report scatter near the $0.1$--$0.3$ dex level \citep{vandesande2015, muzzin2009, battisti2024}, although their samples and labels differ from ours. A single F150W point cannot support conventional photometric SED fitting, and NIRSpec slit losses recover only $24$--$63\%$ of the slitless flux for extended sources \citep{dalmasso2025}. Within these limits, our joint representation reaches a comparable $\sigma_{\mathrm{NMAD}} \simeq 0.24$~dex, whereas the cited studies used multi-band photometry or broader spectroscopic coverage. 
In addition to stellar mass, we report probes of S\'ersic morphology, redshift, and F150W photometry on the frozen embeddings (Appendices~\ref{app:bench} and~\ref{app:breakdown}).

\section{Conclusion and Outlook} \label{sec:conclusion}

Self-supervised models trained on \emph{JWST} data recover physical information from a sparse setup of single-band F150W imaging and the observed-frame $1$--$2~\mu$m PRISM window, reaching $R^2 \sim 0.9$ and $\sigma_{\mathrm{NMAD}} = 0.24$~dex at $1<z\le3$ (within the $0.1$--$0.3$~dex scatter of multi-band SED-fitting studies, though sample and label differences preclude direct comparison). CLIP alignment further raises single-modality performance by $\Delta R^2=+0.04$ (imaging) and $+0.09$ (spectroscopy), narrowing the gap to joint-modality inference.


Several limitations remain. Table~\ref{tab:mass} is confined to $1<z\le3$, while the performance degrades significantly at higher redshifts (Figure~\ref{fig:mass}): we adopt a $1$--$2~\mu$m window to emulate \emph{Euclid} / \emph{Roman} spectral coverage, which at z>3.5 samples the rest-frame ultraviolet and is devoid of features correlated with stellar mass, imposing a \underline{physical limitation}. 
Future work will use the full spectral coverage and improved labels to extend this method to higher redshifts. In addition, the stellar-mass labels derived from one SED fitting algorithm may inherit its biases, and the PRISM slit geometry and position angle encode unmodeled spatial information.

Self-supervised representations thus enable physical parameter estimation under sparse wavelength coverage and should extend to broader regimes, from fewer bands to heterogeneous photometric and spectroscopic combinations. As \emph{JWST} and \emph{Roman} deliver larger heterogeneous surveys, self-supervised alignment of imaging and spectroscopy makes stellar-mass estimation feasible for the growing population of sources that have a single photometric band and a low-resolution spectrum, the data that wide-field parallel observations routinely produce.


\begin{ack}
The authors would like to thank Fred Hamprecht, Josh Speagle, and Mike Smith for valuable discussions and feedback. The authors also  acknowledge institutional support from the Max Planck Institute for Astronomy (MPIA), Heidelberg, and Heidelberg University, Department of Physics and Astronomy. This research was performed on the HPC systems Raven, Viper and Astronodes at the Max Planck Computing and Data Facility (MPCDF). Y.~C.\ acknowledges travel fund support from the European Space Agency (ESA) Directorate of Science to attend the Language AI in the Space Sciences Workshop at Space Telescope Science Institute (STScI) in Baltimore.

This work is based on observations made with the NASA/ESA/CSA \emph{James Webb Space Telescope}. The data were obtained from the Mikulski Archive for Space Telescopes (MAST) at the Space Telescope Science Institute, which is operated by the Association of Universities for Research in Astronomy, Inc., under NASA contract NAS~5-03127 for \emph{JWST}. These observations are associated with programs GO~3383 and GO~4681 (OutThere); GO~1727 (COSMOS-Web); ERS~1345 and DDT~2750 (CEERS); 1176, 1180, 1181, 1210, 1264, 1283, 1286, 1287, 1895, 1963, 2079, 2198, 2514, 2516, 2674, 3215, 3577, 3990, 4540, 4762, 5398, 5997, 6434, 6511, and 6541 (JADES DR5); and the public NIRSpec PRISM programs compiled in the DJA. The authors acknowledge the teams of programs 1345 (PI: S.~Finkelstein), 1727 (PIs: J.~Kartaltepe \& C.~Casey), 2750 (PI: P.~Arrabal Haro), 1895, 1963, 2079, 2514, 3215, 3577, 3990, 6434, and 6541 for developing their observing programs with a zero-exclusive-access period.

The specific data analyzed in this work can be accessed at MAST via \url{https://doi.org/10.17909/2sx3-ad32} (COSMOS-Web \citep{franco2026}), \url{https://doi.org/10.17909/8tdj-8n28} (JADES DR5), and \url{https://doi.org/10.17909/z7p0-8481} (CEERS). Some of the data products presented herein were retrieved from the Dawn \emph{JWST} Archive (DJA; \url{https://dawn-cph.github.io/dja/}). DJA is an initiative of the Cosmic Dawn Center (DAWN), which is funded by the Danish National Research Foundation under grant DNRF140.

\paragraph{Software:} astropy \citep{2013A&A...558A..33A,2018AJ....156..123A,2022ApJ...935..167A},
Python \citep{python}, NumPy \citep{numpy}, PyTorch \citep{pytorch}, PyTorch Lightning \citep{lightning}, grizli \citep{grizli}, msaexp \citep{msaexp}, reproject \citep{reproject}, h5py \citep{h5py}, Optuna \citep{optuna}.

\textbf{AI disclosure:} AI tools (\texttt{GPT-5.6 Sol, GPT-5.5, GPT-5.4, GPT-5.3-codex}, OpenAI, and \texttt{Opus 4., Opus 5, Opus 5.5, Haiku 4.5}, Anthropic) were used to assist with scripts for batch retrieval of survey data, FSDP training scheduling and monitoring, and debugging of the processing workflow. Limited AI assistance was also used for grammar and formatting in this manuscript. No AI tools were used to develop the model's core components or to write any section of the manuscript. All AI-assisted suggestions were reviewed, tested where applicable, and adopted by the authors, who remain fully responsible for the scientific analysis, results, conclusions, and this manuscript.

\end{ack}

{\small
\printbibliography
}


\appendix

\section{Data Preparation and Pre-processing} \label{app:data}

This appendix assembles the paired \emph{JWST} imaging and NIRSpec spectra of individual galaxies used in this work. Each source must be linked across its two observing modes before training begins. The steps below cover survey selection and imaging cutouts, spectral extraction with quality filtering, and the positional cross-matching that ties the modes together.

\subsection{Imaging} \label{app:image}
We construct \emph{JWST}/F150W imaging samples from the NIRISS pure-parallel survey OutThere and the NIRCam surveys COSMOS-Web, JADES, and CEERS, which together span a wide range of depth and area coverage. \textbf{OutThere} is a Cycle~2 \emph{JWST}/NIRISS pure-parallel survey (GO~3383 and GO~4681), with $91$ fields over a nominal $626.6~\mathrm{arcmin}^{2}$; we use F150W direct imaging from the $41$ fields, covering approximately $174~\mathrm{arcmin}^{2}$, with an integration time of $0.08$--$2.86$~hours per pointing (median $0.24$~hr). We uniformly reduce the mosaics with \texttt{grizli} \citep{grizli} and reproject them from the grizli output grid of $0.04''$ to the common $0.03''$ pixel scale. \textbf{COSMOS-Web} (PID~1727) is a Cycle~1 $255$-hour NIRCam Treasury program covering $0.54~\mathrm{deg}^{2}$ in F115W, F150W, F277W, and F444W \citep{cosmos}; we use the public F150W mosaics at $0.03''$ per pixel \citep{franco2026} with an exposure time spanning $0.07$--$0.286$~hr. \textbf{JADES} is a $770$-hour NIRCam and NIRSpec guaranteed-time program in GOODS fields \citep{jades}; we use the F150W mosaics from JADES DR5 \citep{johnson2026}, which combine JADES imaging with uniformly reduced public archival observations, spanning approximately a $42~\mathrm{arcmin}^{2}$ deep region and a $167~\mathrm{arcmin}^{2}$ medium region. \textbf{CEERS} (PID~1345) is a $77.2$-hour Early Release Science program covering approximately $90~\mathrm{arcmin}^{2}$ in the Extended Groth Strip \citep{ceers}; we use its public NIRCam F150W mosaics at $0.03''$ per pixel \citep{bagley2023}, with typical integrations of approximately $0.8$~hr per pointing.

\begin{table}[ht]
\centering
\caption{Dataset overview. All counts are after data pre-selection. The entire dataset used in our work contains multi-survey F150W imaging and NIRSpec PRISM/CLEAR spectra from the DJA archive. The last column reports matches between imaging and spectra within $0.3''$: the imaging rows summarize all matched cutouts, while the DJA row counts the number of unique matched spectra. The imaging total exceeds the spectroscopic total since a spectrum can be selected by more than one cutout, either where different survey footprints overlap or where one survey yields two neighboring detections of the same source.}
\label{tab:data}
\begin{tabular}{lllcc}
\toprule
Survey & Program & Instrument & \# samples & \# matched cutouts \\
\midrule
\href{https://github.com/OutThere-JWST/NISPureParallel}{OutThere}
            & GO~3383/4681 & NIRISS & $143{,}296$ & $474$ \\
\href{https://cosmos.astro.caltech.edu/page/cosmosweb}{COSMOS-Web}
            & PID 1727     & NIRCam & $749{,}154$ & $2{,}298$ \\
\href{https://jades-survey.github.io/}{JADES}
            & GTO          & NIRCam & $409{,}124$ & $5{,}999$ \\
\href{https://ceers.github.io/}{CEERS}
            & PID 1345     & NIRCam & $77{,}589$  & $2{,}893$ \\
\cmidrule(lr){1-5}
\textbf{Imaging} & --- & --- & $\mathbf{1{,}379{,}163}$ & $\mathbf{11{,}664}$ \\
\midrule
\textbf{Spectra (\href{https://dawn-cph.github.io/dja/}{DJA})} & multi-program & NIRSpec & $25{,}025$ & $\mathbf{10{,}925}$ \\
\bottomrule
\end{tabular}
\end{table}

The four surveys use different flag vocabularies and signal-to-noise definitions; the resulting source densities vary across surveys.

\textbf{Image selection and pre-processing:} Sources are drawn from each survey's own detection catalog, and we apply two requirements for pre-selection: a quality or bad-region flag cut, which removes sources contaminated by imaging artifacts, and a positive integrated signal-to-noise ratio, which removes detections with negative flux and low S/N. Specifically, COSMOS-Web uses $\texttt{warn\_flag} \le 3$ together with $\texttt{snr\_f150w} > 0$ from the \texttt{COSMOSWeb\_mastercatalog\_v1\_photom} catalog \citep{shuntov2025}; JADES adopts a source quality flag $\texttt{F150W\_FLAG} = 0$ from the DR5 photometric catalog \citep{robertson2026}, together with the Kron total-flux significance $\texttt{F150W\_KRON\_S}/\texttt{F150W\_KRON\_S\_e} > 0$; OutThere relies on the extraction flag $\texttt{FLAG} < 4$ on the per-field IR detection catalog together with the detection significance $\texttt{FLUX\_AUTO}/\texttt{FLUXERR\_AUTO} > 0$. CEERS applies the bad region flag $\texttt{BAD\_REGION\_FLAG} = 0$ from \texttt{ceers\_cat\_v1.0} \citep{cox2026}, and $\texttt{F150W\_FLUX}/\texttt{F150W\_FLUXERR\_EMP} > 0$. After pre-selection, we extract source-centered cutouts from the corresponding mosaics of size $128\times128$ pixels with a common $30$~mas pixel scale and a shared $\mathrm{MJy\,sr^{-1}}$ unit, so that the four surveys share a uniform model input. We note the cutout size is larger than the input size of the image encoder ($100\times100$~px), as the extra pixels are used for the random rotation and flip to ensure no black corners are included in the augmentations.

\subsection{Spectra} \label{app:spectra}
We construct the spectroscopic sample from the NIRSpec/PRISM observations in the DAWN \emph{JWST} Archive. The DJA provides uniform reductions from multiple observational programs using \texttt{msaexp} \citep{msaexp}; details of the archive's reduction are given in \citet{Heintz2024, degraff2025}. The archive samples all spectra onto one shared $473$-pixel observed-frame wavelength grid for PRISM observations. We extract only the $1$--$2~\mu$m window, which corresponds to a fixed grid of $56$ pixels across all spectra.

\textbf{Spectra selection and pre-processing:} we apply two selections. We keep spectra with a DJA visual-inspection quality grade $>0$ (no clearly visible data corruption), and we require at least $50\%$ of the $56$ pixels inside the $1$--$2~\mu$m window to be valid pixels to remove bad-quality or half-empty spectra. We apply no signal-to-noise or redshift cut at this stage. The archive stores fluxes as $f_\nu$ in $\mu$Jy; we convert them to $f_\lambda \propto f_\nu/\lambda^{2}$, and we take the per-pixel uncertainties from the \texttt{full\_err} column and convert them alongside the flux.

\subsection{Source Cross-matching} \label{app:match}
We match each imaging survey independently against the $25{,}025$ spectra that pass the selection in Appendix~\ref{app:spectra}. The spectroscopic archive repeats many observations from different pointings, so the match must pick the PRISM pointing closest to each image center. We link every image-detection position to its nearest DJA neighbor within $0.3''$, which yields $11{,}664$ matched cutouts for $10{,}925$ unique spectra.

\textbf{Duplicated observations:} More than one cutout matches $716$ unique spectra. About half of these, $353$, arise where the NIRISS and NIRCam imaging footprints overlap and both instruments image the same source ($227$ spectra shared between OutThere and JADES, $126$ between OutThere and COSMOS-Web); a further $342$ are two neighboring detections within the same survey ($214$ in JADES, $114$ in CEERS, $10$ in COSMOS-Web, $4$ in OutThere), and three or four cutouts match the remaining $21$. We handle such duplicated observations differently in two stages of this work: the self-supervised pre-training of the image and spectrum encoders requires no treatment here since no labels are involved, and the repeated image input is a natural augmentation. For the CLIP alignment and every downstream probe, however, we split our dataset by unique DJA source identifier to create a train/test set so that the repeated cutouts belonging to one spectrum always land on the same side of the split without data leakage.

\section{Model Training Setup} \label{app:model}

\subsection{Image Encoder} \label{app:imagemodel}
\label{app:model_img}

\textbf{Data augmentation and tokenization.} DINOv2 trains by matching two distorted views of the same source in the embedding space, so the augmentation pipeline defines the invariances the encoder learns. We construct a physically motivated augmentation pipeline that mimics the observational invariances a galaxy cutout should preserve. In our standard DINOv2 setting, each cutout yields two \emph{global} crops ($72\times72$~px) and $n_{\mathrm{local}}=8$ \emph{local} crops ($36\times36$~px); the operations forming each crop are, in order:
\begin{enumerate}
    \item \textbf{Random rotation and flips.} Each view is rotated by a random angle from $[-45^\circ,+45^\circ]$ with probability $0.5$; the rotated view is then center-cropped to $100\times100$~px to discard the empty corners that rotation introduces. We next apply random horizontal and vertical flips, each with probability $0.5$, before drawing the random global and local crops described below. These geometric steps together make the embedding invariant to the arbitrary sky orientation of a galaxy.
    \item \textbf{Random crop} to the target size, taken at a location drawn uniformly inside the rotated cutout: $72\times72$~px for a global view and $36\times36$~px for a local one. The random crop encourages the model to align its embedding under different translations and also to distinguish the center object of the cutout from potential neighboring objects. The global and local crop sizes are chosen manually and were found to work stably in preliminary experiments.
    \item \textbf{Gaussian sky noise.} Zero-mean noise is added in the raw-flux domain to simulate a shallower observation, $\sigma_{\mathrm{add}} = s\,\sigma_{\mathrm{sky}}$ as an extra sky noise level, with $s  = |\mathcal{N}(0,w)|$, $w=1$, capped at $s_{\max}=2$. This half-normal distribution has its mode at zero and a wide tail, so most views receive almost no extra noise. $\sigma_{\mathrm{sky}}$ is the background level measured in the cutout's own survey pointing, so the augmentation is scaled by the pointing-specific noise level. Noise is applied with probability $1.0$, $0.1$, and $0.5$ to the first global, second global, and local views respectively, taken from the original DINOv2 implementation.
    \item \textbf{Intensity stretch.} Raw pixel fluxes span a high dynamic range, so each augmented view is passed through an $\mathrm{arcsinh}$ stretch, following \citet{stein2021, astroclip2023, platonicuniverse}. 
    This stabilizes ViT training and prevents the brightest pixels from dominating the learned representation.
\end{enumerate}

\textbf{Pre-training setup.}
We pre-train on the complete multi-survey F150W sample of Table~\ref{tab:data}, split per survey into $90\%/10\%$ train/validation sets. We draw each batch at random from the pooled data across all surveys, so every iteration reflects the relative object distribution of the full dataset, and a fixed seed of $42$ is used for both the data randomization and the model initialization.

We adopt the standard vision transformer from the DINOv2 implementation \citep{dosovitskiy2021} as the backbone for both teacher and student networks, and the AdamW optimizer \citep{adamw} for updating the student parameters. In general, DINOv2 learns by self-distillation: the teacher and the student see different augmented views of the same source, and the student is trained to reproduce the teacher's output so that the representation becomes invariant to the transformations described above. The teacher is never updated by gradient descent but follows a slow exponential moving average of the student, which supplies a stable target and prevents the two networks from collapsing onto a constant. Two objectives are introduced: the self-DIstillation with no labels (DINO) loss matches the projected \texttt{[CLS]} token between the student and teacher, while an iBOT loss (Image BERT Pre-Training with Online Tokenizer \citep{ibot,oquab2024}) randomly masks part of the student's tokens and asks it to recover the teacher's tokens. A third term, the KoLeo regularizer, is a small penalty that keeps the embeddings within a batch from clustering too tightly. The detailed mathematical formulation of the three terms is given in \citet{oquab2024}.

Values in bold in Table~\ref{tab:image_config} are set manually and depart from the default DINOv2 configuration \citep{oquab2024}. \textbf{Model and training scale:} our training set ($1.4\times10^{6}$ cutouts) is two orders of magnitude smaller than the LVD-142M set of DINOv2, so we narrow the backbone and the projection head and lower the drop-path rate accordingly; given our hardware and compute budget, we also set the batch size and the numbers of iterations and epochs for a correspondingly smaller training run. \textbf{Augmentation settings:} for the tokenization and augmentation, we use overlapping patches of size $6$ and stride $3$ to increase the spatial resolution of the tokens, and we lower the maximum iBOT mask ratio from $0.5$ to $0.3$, as most of each cutout is background sky. In preliminary runs, this setup gave stable training and improved the performance of a morphology probe, so we adopted it without further tuning. \textbf{Optimization:} the default optimization settings led to unstable training in preliminary runs, so we adjust the optimizer hyperparameters, guided by previous self-supervised models of galaxy images \citep{stein2021, astroclip2023}: we lower the learning rate and weight decay, and use a lower KoLeo weight, a higher teacher momentum, a shorter teacher-temperature warm-up from a higher starting value, and a longer freeze of the last layer, which together yield a stably converging training loss.

The training runs in FSDP-sharded \texttt{bf16} mixed precision, at a per-GPU batch size of $56$ over $20$ A100 GPUs. Training takes approximately $40$ hours in the configuration described in this section, and we use the teacher weights from the final epoch for the CLIP alignment and all downstream probes.

\begin{table}[ht]
\centering
\small
\caption{Adopted image-encoder pre-training configuration. Values in \textbf{bold} are the ones we set for this dataset rather than inheriting from the standard DINOv2 recipe, as described in Appendix~\ref{app:imagemodel}.}
\label{tab:image_config}
\begin{tabular}{lll}
\toprule
Group & Hyperparameter & Value \\
\midrule
\multirow{5}{*}{Backbone}
 & embedding dim.\ / depth / heads      & $\mathbf{512}$ / $12$ / $\mathbf{8}$ \\
 & MLP ratio / input channels           & $4$ / $\mathbf{1}$ \\
 & register tokens                      & $7$ \citep{darcet2024} \\
 & drop-path rate                       & $\mathbf{0.2}$ \\
 & LayerScale init                      & $10^{-5}$ \\
\midrule
\multirow{2}{*}{Projection head}
 & prototypes / bottleneck              & $\mathbf{16{,}384}$ / $\mathbf{192}$ \\
 & layers / hidden dim.                 & $3$ / $\mathbf{768}$ \\
\midrule
\multirow{6}{*}{Views \& tokenization}
 & global / local crop size             & $\mathbf{72}$ / $\mathbf{36}$ px \\
 & number of global / local crops                & $2$ / $8$ \\
 & center crop after rotation           & $\mathbf{100}$ px \\
 & patch size / stride                  & $\mathbf{6}$ / $\mathbf{3}$ \\
 & noise scale $w$ / cap $s_{\max}$     & $\mathbf{1.0}$ / $\mathbf{2.0}$ \\
 & iBOT mask ratio / sample prob.       & $\mathbf{[0.1,\,0.3]}$ / $0.5$ \\
\midrule
\multirow{3}{*}{Objectives}
 & DINO / iBOT / KoLeo weight           & $1.0$ / $1.0$ / $\mathbf{0.03}$ \\
 & teacher momentum (start $\to$ end)   & $\mathbf{0.994} \to 1.0$ \\
 & teacher temp.\ (warm-up epochs) & $\mathbf{0.06} \to 0.07$ over $\mathbf{10}$ ep. \\
\midrule
\multirow{7}{*}{Optimization}
 & optimizer / $\beta$                  & AdamW / $(0.9,\,0.999)$ \\
 & base learning rate                   & $\mathbf{8\times10^{-5}}$ at batch $1024$ \\
 & min.\ learning rate                  & $10^{-6}$ \\
 & weight decay (start $\to$ end)       & $\mathbf{0.004} \to \mathbf{0.04}$ \\
 & warm-up epochs / freeze last layer   & $10$ / $\mathbf{3}$ \\
 & gradient clipping                    & $3.0$ \\
 & layer-wise decay / patch-embed mult. & $0.9$ / $0.2$ \\
\midrule
\multirow{3}{*}{Schedule \& infra.}
 & epochs $\times$ iterations           & $\mathbf{300} \times \mathbf{1000}$ \\
 & batch size                           & $\mathbf{56}$ per GPU $\times\,20$ GPUs \\
 & precision                            & FSDP \textbf{\texttt{bf16}} mixed \\
\bottomrule
\end{tabular}
\end{table}

\subsection{Spectrum Encoder}
\label{app:model_spec}

\textbf{Preprocessing and tokenization.} Each spectrum is first passed through an $\mathrm{arcsinh}$ stretch normalized by the median absolute flux of that spectrum, $f'_i=\operatorname{arcsinh}(f_i/a)$ with $a=\operatorname{median}_{i\in V}|f_i|$, and the stretched spectrum is then standardized to zero mean and unit variance, with the mean and standard deviation measured on the arcsinh-stretched flux before standardization. Similar to the imaging case, this pre-processing step compresses the high dynamic range and removes the absolute scale of the data for stable training. We then divide the standardized spectrum into overlapping patch tokens of $P=4$ pixels with stride $2$. Each token is further prepended with the mean and standard deviation of the stretched spectrum, and combined with a Fourier positional encoding of its mean observed-frame wavelength, following \citet{spectral_token,omnispectra2026,aion2025}. A token is marked as invalid and excluded from attention and from the loss, if any of the token pixels are flagged as bad in the DJA dataset.

\textbf{Masked reconstruction.} Pre-training follows the masked-modeling recipe of \citet{he_masked_2021}: a fraction of the valid tokens is hidden from the encoder and reconstructed from the surrounding context. Tokens are randomly hidden in contiguous blocks of two tokens, separated by at least one visible token. Because consecutive patches overlap by half of their pixels, contiguous masking prevents data leakage from adjacent patches. Eight layers of transformer blocks with residual connections encode the input token sequence, while the marked invalid tokens are excluded end to end. A single linear decoder finally maps the latent tokens back to the patch pixels. As described in Section~\ref{sec:method}, the loss is weighted by the reported uncertainties. The per-pixel errors are propagated through the same stretch and standardization as the flux, $\epsilon'_i=\epsilon_i/(a\sqrt{1+(f_i/a)^2}\,\sigma_{f'})$. The token variance $\sigma_t^2$ is the mean squared propagated uncertainty over its valid pixels, and each masked token receives $w_t=1/\max(\sigma_t^2,\sigma^2_{\min})$. The floor $\sigma^2_{\min}$ prevents a handful of very clean tokens from dominating the gradient, and tokens without a usable uncertainty are given zero weight.

\textbf{Hyperparameter search and training setup.} We adopt a transformer masked autoencoder because this design has been widely adopted and well tested for astronomical spectra \citep{astroclip2023, spectral_token, omnispectra2026, aion2025}. Unlike the image encoder, whose configuration was adapted by hand from a well-tested reference recipe, the spectrum encoder has no established configuration at this resolution and sample size. We use \textit{Optuna}, a hyperparameter optimizer with its Tree-structured Parzen Estimator \citep{optuna}, to search over promising configurations. The search objective is the $\sigma_{\mathrm{NMAD}}$ of a linear redshift probe, as introduced in Section~\ref{sec:results}. Each search trial trains for $300$ epochs and fits the linear probe on redshift every $20$ epochs, averaged over three random seeds, and the best $\sigma_{\mathrm{NMAD}}$ reached during the run is taken as the trial score. The joint search covers the model width and depth, the head dimension, dropout, the mask ratio, the variance floor $\sigma^2_{\min}$, the learning rate, and the weight decay, over $180$ trials of pre-training in total. The patch size and stride are held fixed at $P=4$ and $S=2$, which provided stable training in early experiments of this work. The searched ranges and the final \textit{Optuna} configuration are listed in Table~\ref{tab:spectrum_config}. The search is run on a single H100 GPU and takes approximately $15$ hours in total.

The final model adopts this configuration unchanged, as shown in the right column of Table~\ref{tab:spectrum_config}. For Optuna and pre-training, we adopt a $90\%/10\%$ train/validation split of the DJA sample. We train the final model with the same machine setup as the parameter search, for approximately $30$ minutes. The checkpoint with the lowest validation reconstruction loss on the masked tokens is used for the CLIP alignment and all downstream probes. Unlike the image teacher, whose exponential-moving-average weights converge smoothly to the final epoch, we take the spectrum checkpoint at its lowest validation reconstruction loss, because the masked-reconstruction objective can overfit in the later epochs at this sample size.

\subsection{CLIP alignment}\label{app:model_clip}

Following \citet{buck,astroclip2023,clip}, we use contrastive learning to align the two pre-trained encoders in a shared latent space. The physical motivation is that imaging and spectroscopy describe the same galaxy through complementary views. Pulling matched embedding vectors together and pushing unmatched pairs apart, the alignment emphasizes the physical structure shared by both modalities and makes their representations comparable. At inference, a single observed modality can stand in for the missing one.

\textbf{Attention pooling and scale re-injection.} The alignment operates on one vector per modality, so we first compress each encoder output into a single embedding in two steps. We freeze both encoders and collapse their token sequences via attention pooling \citep{lee2019}, in which a single learned query attends over a sequence and returns the attention-weighted sum of its tokens as one fixed-length vector. For modality $m$ with encoder tokens $x_{m,i}$, the pooled representation is

$$
    z_m=\sum_i \alpha_{m,i}W_{v,m}x_{m,i}, \qquad
    \alpha_{m,i}=\operatorname{softmax}_i\!\left(\frac{q_m^{\top}W_{k,m}x_{m,i}}{\sqrt{d_m}}\right).
$$

The query attends to the entire token sequence of one encoder, with the invalid tokens masked out in case of the spectrum encoder. The image branch assumes all patch tokens are valid, because the imaging cutouts are pre-filtered for invalid pixels (Appendix~\ref{app:image}); only the spectrum branch requires explicit masking. We keep the attention head configuration the same as in the pre-trained model, with eight heads for the image and four for the spectrum, so the pooling reuses each encoder's head count. In the second step, as per-object normalization removes the absolute flux scale for both modalities, we add it back by linearly projecting the arcsinh-stretched mean and standard deviation $s_m$, $\tilde z_m=z_m+W_{s,m}\operatorname{arcsinh}(s_m/s_{m,\mathrm{ref}})$, where $s_{m,\mathrm{ref}}$ is the fixed per-column median of $|s_m|$. We concatenate the same mean and standard deviation to the unaligned embeddings in the downstream tasks, so the probes read identical information before and after alignment.

\textbf{CLIP training.} From the two pooled, scale-corrected vectors above, one small MLP per modality, with hidden widths $(512,\,256)$ for the image and $(128,\,256)$ for the spectrum, projects the embeddings into $L_2$-normalized, $256$-dimensional vectors, $u_i$ and $v_i$. For a batch of $B$ matched pairs, we then align the embedding pair inside a training batch with a contrastive loss \citep{oord2018,clip},

$$
    \mathcal{L}_{\mathrm{CLIP}}
    =-\frac{1}{2B}\sum_{i=1}^{B}\left[
    \log\frac{\exp(u_i^{\top}v_i/\tau)}{\sum_j\exp(u_i^{\top}v_j/\tau)}
    +\log\frac{\exp(v_i^{\top}u_i/\tau)}{\sum_j\exp(v_i^{\top}u_j/\tau)}
    \right].
$$

The alignment head, comprising the attention pooling, the scale projection, and the two MLPs, holds $1.7$M trainable parameters. We train it on the full $11{,}664$ pairs ($10{,}925$ unique spectra) with a $90\%/10\%$ train/validation split grouped by unique DJA source identifier. We use a fixed temperature $\tau=0.1$, batch size $B=512$, and a standard AdamW optimizer with weight decay $0.04$. The learning rate warms up linearly to $10^{-3}$ over $1000$ iterations and then decays with a cosine schedule to $10^{-6}$. Training uses \texttt{bf16} mixed precision on a single H100 GPU and completes in $\sim 30$ minutes.

\begin{table}[ht]
\centering
\small
\caption{Spectrum encoder pre-training configuration and parameter search space with \textit{Optuna}. Entries specified in the third column were optimized jointly with Optuna's TPE sampler; \texttt{fixed} rows were held constant. The last column gives the best \textit{Optuna} configuration that is adopted for the final training.}
\label{tab:spectrum_config}
\begin{tabular}{lllc}
\toprule
Group & Hyperparameter & Search space / Fixed & Final \\
\midrule
\multirow{5}{*}{Backbone}
 & embedding dim.\ $d$   & $\{32,\ 64,\ 128,\ 192,\ 256\}$ & $128$ \\
 & transformer layers    & $\{4,\ 6,\ 8\}$        & $8$ \\
 & attention heads       & $d\,/\,\{8,\ 16,\ 32\}$      & $4$ \\
 & FFN ratio             & \texttt{fixed}              & $4.0$ \\
 & dropout               & $[0.0,\ 0.3]$               & $0.05$ \\
\midrule
\multirow{8}{*}{Masking / objective}
 & patch size $P$ / stride $S$ & \texttt{fixed}        & $4$ / $2$ \\
 & invalid-pixel threshold & \texttt{fixed}            & $25\%$ \\
 & continuous block length & $\{1,\ 2,\ 3,\ 4\}$       & $2$ \\
 & mask ratio            & $[0.30,\ 0.75]$             & $0.60$ \\
 & min.\ visible tokens  & \texttt{fixed}              & $4$ \\
 & loss weighting        & \texttt{fixed}              & inv.\ var. \\
 & $\sigma^2_{\min}$     & $[0.01,\ 2.0]$ log-uniform  & $0.10$ \\
 & decoder head          & \texttt{fixed}              & linear \\
\midrule
\multirow{6}{*}{Optimization}
 & optimizer / $\beta$   & \texttt{fixed}              & AdamW / $(0.9,\ 0.95)$ \\
 & learning rate         & $[10^{-4},\ 10^{-3}]$ log-uniform & $2\times10^{-4}$ \\
 & weight decay          & $[10^{-5},\ 10^{-1}]$ log-uniform & $4\times10^{-3}$ \\
 & warm-up steps         & $[200,\ 2000]$ log-uniform  & $1000$ \\
 & min.\ learning rate   & \texttt{fixed}              & $10^{-6}$ \\
 & batch size            & \texttt{fixed}              & $1024$ \\
\midrule
\multirow{2}{*}{Schedule \& infra.}
 & epochs $\times$ iterations & \texttt{fixed}         & $300 \times 21$ \\
 & precision / devices   & \texttt{fixed}              & \texttt{bf16} mixed / $1$ GPU \\
\bottomrule
\end{tabular}
\end{table}

\section{Single-Encoder Benchmarks}
\label{app:bench}

Before the two modalities are aligned, each encoder is evaluated on its own, against a property that only that modality can constrain: morphology for the image encoder and redshift for the spectrum encoder. Both are read out from the frozen encoder, so any performance is attributable to the pre-training rather than to the capacity of the probe.

\textbf{Seed sensitivity.} The linear-probe metrics are reported as mean and standard deviation over $40$ random train/test splits; the split-to-split scatter is far smaller than the differences between the representations compared in Table~\ref{tab:mass}. For the stellar-mass probe of Table~\ref{tab:mass}, the split-to-split standard deviations are at most $0.011$ for $R^2$, $0.009$~dex for $\sigma_{\mathrm{NMAD}}$, and $0.7\%$ for the outlier fraction. We therefore treat pre-training as fixed and attribute the reported scatter to the probe's random splits rather than to pre-training sensitivity, since both encoders are trained with label-free self-supervised objectives on large datasets and strong augmentations.

\subsection{Qualitative behavior}

\begin{figure}[ht]
    \centering
    \includegraphics[width=0.95\linewidth]{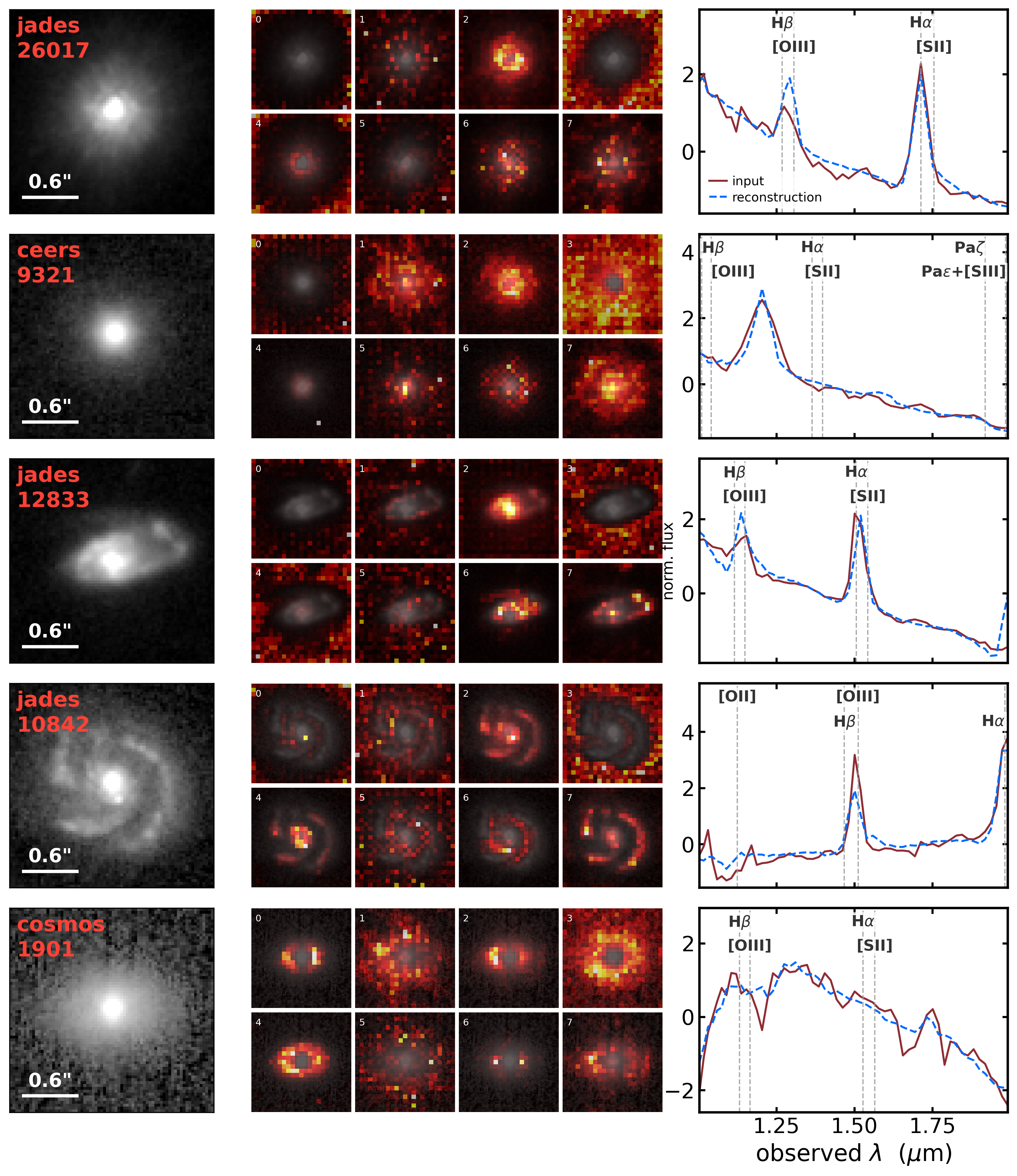}
    \caption{Qualitative behavior of the two frozen encoders on five cross-matched objects, one per row. Left: the F150W cutout given to the image encoder, annotated with the imaging survey and the DJA spectrum identifier, with a $0.6''$ scale bar. Middle: the attention of the \texttt{[CLS]} token to the patch tokens in the final transformer block of the frozen teacher, shown separately for each of the eight heads; the heads specialize on different parts of the source and of the background. Right: the normalized, $\mathrm{arcsinh}$-stretched PRISM spectrum over the observed-frame $1$--$2~\mu$m window (dark red) with the spectrum-encoder reconstruction (blue, dashed), obtained by masking every three consecutive tokens in turn and predicting them from the remaining sequence; vertical lines mark the strong emission lines at the catalog redshift.}
    \label{fig:evaluation}
\end{figure}

Figure~\ref{fig:evaluation} shows how the two frozen encoders respond to the same cross-matched objects. As the image encoder is trained fully self-supervised in the embedding space, its behavior is difficult to assess from the model alone. To build qualitative insight, we follow the DINO \citep{dino1} approach to visualization and check the attention of the \texttt{[CLS]} token to the patch tokens in the final transformer layer, that is the per-token weights $\mathrm{softmax}(qk^{\top}/\sqrt{d})$ carried by the \texttt{[CLS]} query $q$ to all patch token keys $k$ and then resampled onto the cutout pixel grid for display. The eight heads are shown separately: individual heads isolate the bright core, the outskirts, and the background, while in the clumpy and spiral sources, some heads trace the substructure itself, closely following the arms. The heads recover, without supervision, the morphology that an expert would delineate by eye. For the spectrum encoder, we reconstruct the same objects with the masked protocol applied in pre-training (Sections~\ref{sec:method} and~\ref{app:model_spec}). We hide each token and predict it from the rest of the sequence iteratively. In particular, we mask and reconstruct every three consecutive tokens, an even stronger masking than pre-training. We obtain the final reconstructed spectrum by averaging the overlapping pixels between tokens, shown in the right column of Figure~\ref{fig:evaluation}. In general, the reconstruction follows the continuum and recovers the strong emission lines. Although the reconstructed pixels are hidden, the model is able to infer those spectral structures from the rest of the spectrum.

\subsection{Image encoder: S\'ersic parameters} \label{app:sersic}

\begin{figure}[ht]
    \centering
    \includegraphics[width=1\linewidth]{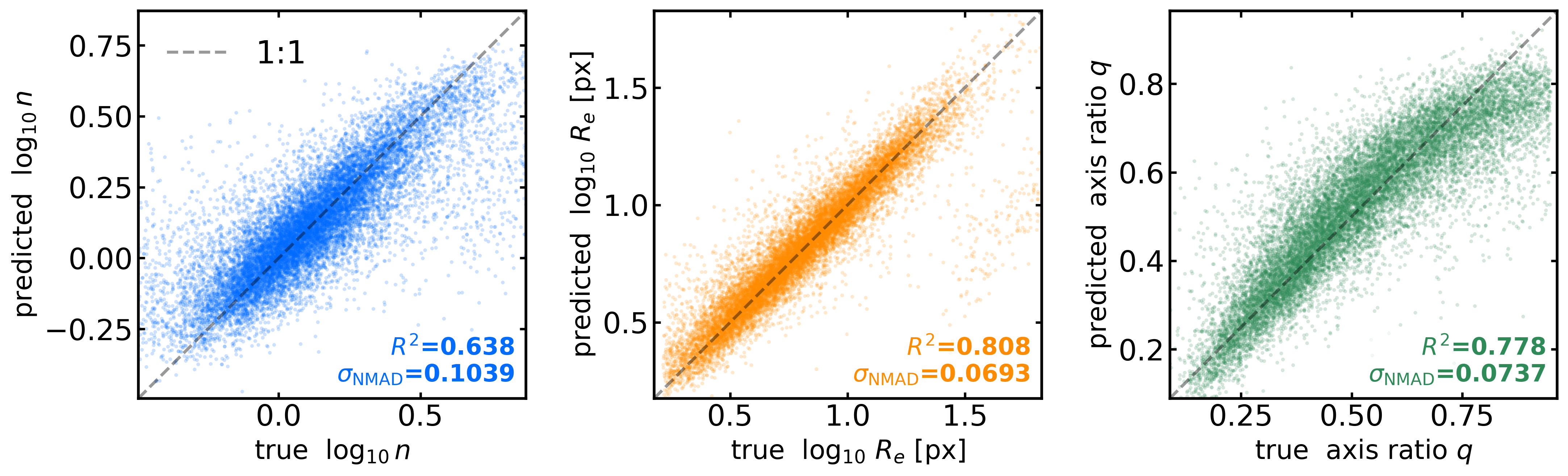}
    \caption{Linear probes of the S\'ersic index (left), effective radius (middle), and axis ratio (right) from the frozen image embedding, for COSMOS-Web sources that are resolved, and detected at $\mathrm{S/N}>30$. Predictions are shown for the test half of a $50\%/50\%$ split; the gray dashed line is the one-to-one relation and $R^2$ and $\sigma_{\mathrm{NMAD}}$ are quoted in each panel. Split-to-split standard deviation changes $R^2$ and $\sigma_{\mathrm{NMAD}}$ by $0.008$ and $0.001$.}
    \label{fig:sersic}
\end{figure}

We fit ridge probes from the frozen embedding to the \texttt{SourceXtractor++} S\'ersic parameters of the COSMOS-Web catalog \citep{shuntov2025}: the index $n$, the effective radius $R_e$, and the axis ratio $q$, with $n$ and $R_e$ probed in $\log_{10}$. We find that catalog fitting uncertainties from faint objects set an irreducible floor on the probe, so we filter the COSMOS-Web dataset using two criteria. Sources that are resolved ($R_e$ above the F150W PSF FWHM, $1.6$~px) and detected at $\mathrm{S/N}>30$ are included. This leaves $134{,}607$ sources, corresponding to $\sim18\%$ of the total sample of COSMOS-Web cutouts. As the number of samples is too large for a single fit, we randomly draw $30{,}000$ subsamples and split $50\%/50\%$ into training and test sets. Similar to the mass probe (Section~\ref{sec:results}), we repeat this sampling process over $40$ different seeds. The results are shown in Figure~\ref{fig:sersic}. The probe recovers $(R^2,\ \sigma_{\mathrm{NMAD}})$ of $(0.65,\ 0.10)$ for $\log_{10} n$, $(0.81,\ 0.07)$ for $\log_{10} R_e$, and $(0.79,\ 0.07)$ for $q$. The S\'ersic index is recovered worst, consistent with $n$ being the parameter whose catalog fits are themselves least well constrained. Lowering the threshold to $\mathrm{S/N}>20$ ($184{,}884$ sources) reduces the three $R^2$ values (for $n$, $R_e$, and $q$) to $0.57$, $0.77$, and $0.74$ while the spread $\sigma_{\mathrm{NMAD}}$ of the targets stays the same, which shows that the difference is due to label noise.

\subsection{Spectrum encoder: redshift} \label{app:redshift}

\begin{figure}[ht]
    \centering
    \includegraphics[width=1\linewidth]{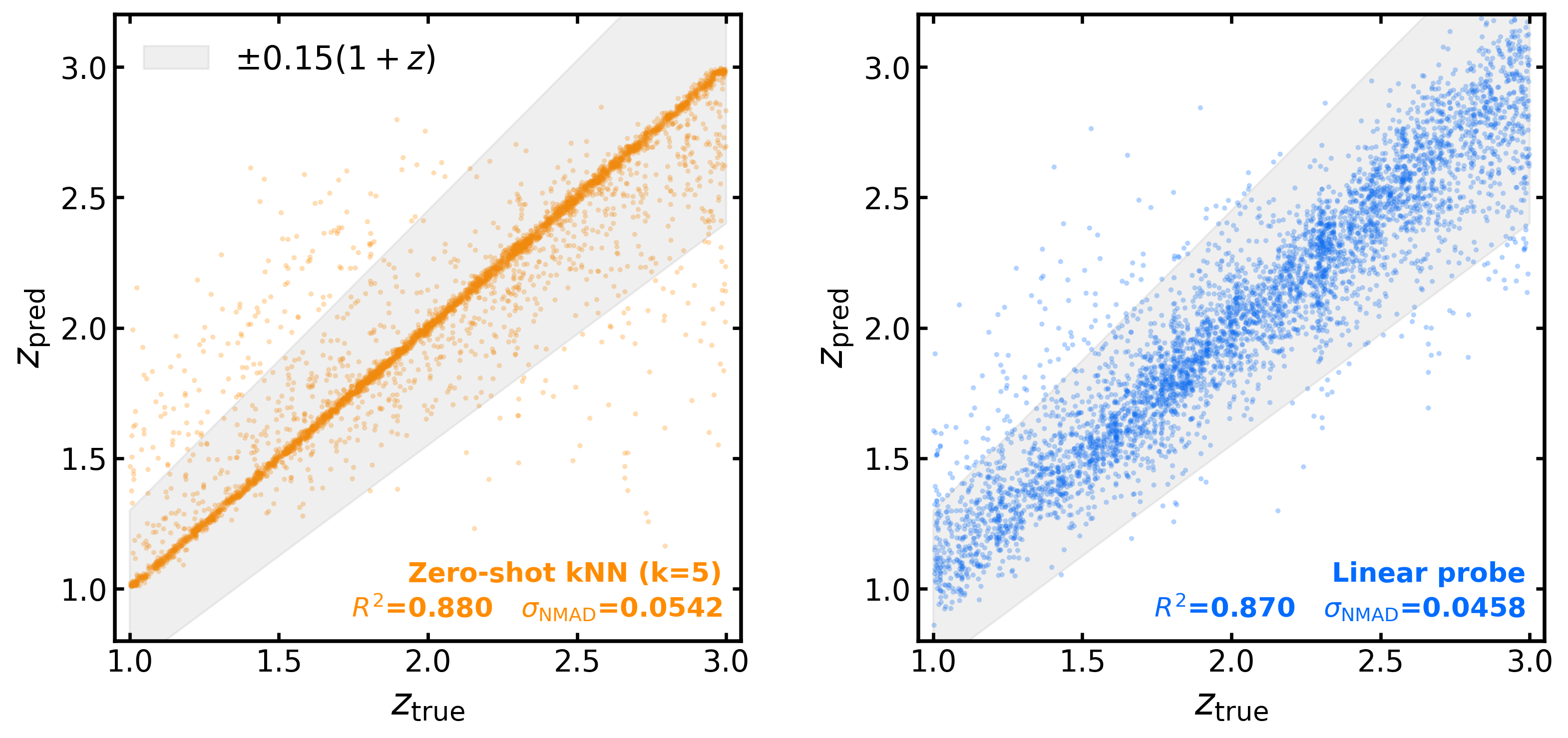}
    \caption{Redshift probes of the frozen spectrum embedding at $1<z\le3$. Left: predicted versus catalog redshift for a non-parametric five-nearest-neighbor probe. Right: the same for a ridge probe. The labeled sample is split equally into a reference/training half and a test half, and every point is a test object; the shaded band marks $|z_{\mathrm{pred}}-z_{\mathrm{cat}}|/(1+z_{\mathrm{cat}})<0.15$. $R^2$ and $\sigma_{\mathrm{NMAD}}$ are quoted in each panel.}
    \label{fig:spectrum_bench}
\end{figure}

The spectrum encoder is evaluated by probing the DJA catalog redshift $z_{\text{spec}}$. Model performance is redshift dependent because the fixed observed-frame $1$--$2~\mu$m window samples a different part of the rest-frame SED at every redshift, so we benchmark in the same $1<z\le3$ interval in which we have shown in Section~\ref{sec:results} that the downstream mass estimate is strongest. We further require more than $90\%$ of the pixels in the window to be valid, a stricter cut than the one used for pre-training, to prevent uneven coverage across the sample. This leaves $9{,}172$ spectra, split into equal train/test sets grouped by DJA object ID. Because the spectrum benchmark requires no image cross-match, it uses all $9{,}172$ spectra, more than the $3{,}452$ cross-matched pairs of the mass and joint probes. The probe results are shown in Figure~\ref{fig:spectrum_bench}. Redshift is recovered with $(R^2,\ \sigma_{\mathrm{NMAD}})=(0.880,\ 0.054)$ for a non-parametric $5$-NN probe and $(0.870,\ 0.046)$ for the ridge probe, and $5.1\%$ and $4.3\%$ of objects respectively with $|\Delta z|/(1+z) \ge 0.15$. That a linear map already reaches $R^2=0.87$ without updating the encoder shows that redshift is close to being linearly decodable from the pre-trained representation. This is consistent with other self-supervised spectral models trained on various datasets \citep{astroclip2023, omnispectra2026, spectral_token}, all of which find redshift to be the property most directly organized in the embedding space. 

We note that redshift is probed again in Appendix~\ref{app:breakdown}, where Table~\ref{tab:targets} reports lower $R^2$ ($0.835$ and $0.816$). The difference arises from the choice of pooling: the probe reads the full token output of the model ($27\ \text{tokens}\times 128 = 3456$-d), while Table~\ref{tab:targets} applies masked mean and max pooling ($258$-d) to obtain an embedding of comparable dimensionality to the image ($512$-d). Since redshift information is highly localized, pooling across tokens dilutes the signal.

\section{Stellar Mass Probe and CLIP Alignment} \label{app:breakdown}

\textbf{The effect of CLIP alignment on linear probes:} Table~\ref{tab:mass} shows that CLIP alignment improves each single modality for stellar mass prediction while leaving the joint embedding unchanged. To test how alignment reshapes each modality, we probe the same six embedding configurations of Table~\ref{tab:mass} (the three learned representations in their unaligned and aligned states, excluding the F150W baseline) on three targets where the two modalities carry unequal information: catalog redshift, $\log F_{150\mathrm{W}}$ flux, and stellar mass, using the same $1<z\le3$ sample and the ridge-regression probe setup. We also adopt the same embedding setup as in Table~\ref{tab:mass}. The unaligned embeddings are the direct encoder outputs: the \texttt{[CLS]} token for the image ($512$) and the masked mean and max over the patch tokens for the spectrum ($2\times128$), each concatenated with the two flux statistics that the alignment head also receives, giving $514$ and $258$ dimensions, respectively, and $772$ for their concatenation. The aligned readouts are the direct output of CLIP alignment head introduced in Section~\ref{app:model_clip}, giving $256$-dimensional attention-pooled and scale-reinjected embedding vectors per modality, and $512$ when concatenated. The results are listed in Table~\ref{tab:targets}.

\begin{table}[ht]
\centering
\caption{Linear probes of three targets $R^2$ on the same six representations as in Table~\ref{tab:mass}, at $1<z\le3$ ($3{,}452$ sources). Entries are the mean $R^2$; split-to-split standard deviations are at most $0.013$. The $\log M_\star$ columns reproduce the $R^2$ columns of Table~\ref{tab:mass}. Arrows mark the $R^2$ gains above $0.03$, each of them several times the split-to-split scatter of the paired difference; no entry of the joint readout reaches that level.}
\label{tab:targets}
\setlength{\tabcolsep}{9pt}
\begin{tabular}{lcccccc}
\toprule
& \multicolumn{2}{c}{redshift $z$} & \multicolumn{2}{c}{$\log F_{150\mathrm{W}}$}
& \multicolumn{2}{c}{$\log M_\star$} \\
\cmidrule(lr){2-3}\cmidrule(lr){4-5}\cmidrule(lr){6-7}
Representation & un. & al. & un. & al. & un. & al. \\
\midrule
Image            & $0.139$ & $0.192$\arrup & $0.964$ & $0.969$ & $0.780$ & $0.818$\arrup \\
Spectrum         & $0.835$ & $0.816$ & $0.739$ & $0.882$\arrup & $0.764$ & $0.856$\arrup \\
Image + spectrum & $0.821$ & $0.808$ & $0.964$ & $0.970$ & $0.896$ & $0.902$ \\
\bottomrule
\end{tabular}
\end{table}

Before alignment, each modality performs strongly where it correlates most with the probe target: the image embedding reaches $R^2=0.964$ on flux but is nearly blind to redshift ($R^2=0.139$); the spectrum resolves the redshift ($0.835$) and recovers the flux only to $0.739$, because the NIRSpec slit samples only a fraction of an extended source (slit loss). The alignment moves each modality toward the other's strength, and only there: imaging gains $\Delta R^2 = +0.053 \pm 0.010$ on redshift, while spectroscopy gains $+0.143 \pm 0.009$ on flux. Neither modality is displaced from what it already decodes well: the spectrum keeps its redshift to within $-0.019 \pm 0.007$ and the image keeps its flux to within $+0.005 \pm 0.002$. The stellar-mass gains quoted in Section~\ref{sec:results}, $+0.038$ for imaging and $+0.092$ for spectroscopy, are also shown here. The joint embedding probe, however, does not benefit from CLIP: it recovers redshift, flux, and stellar mass at the same level before and after the alignment. This is consistent with the information ceiling conjectured in Section~\ref{sec:results}.

\begin{figure}[ht]
    \centering
    \includegraphics[width=1\linewidth]{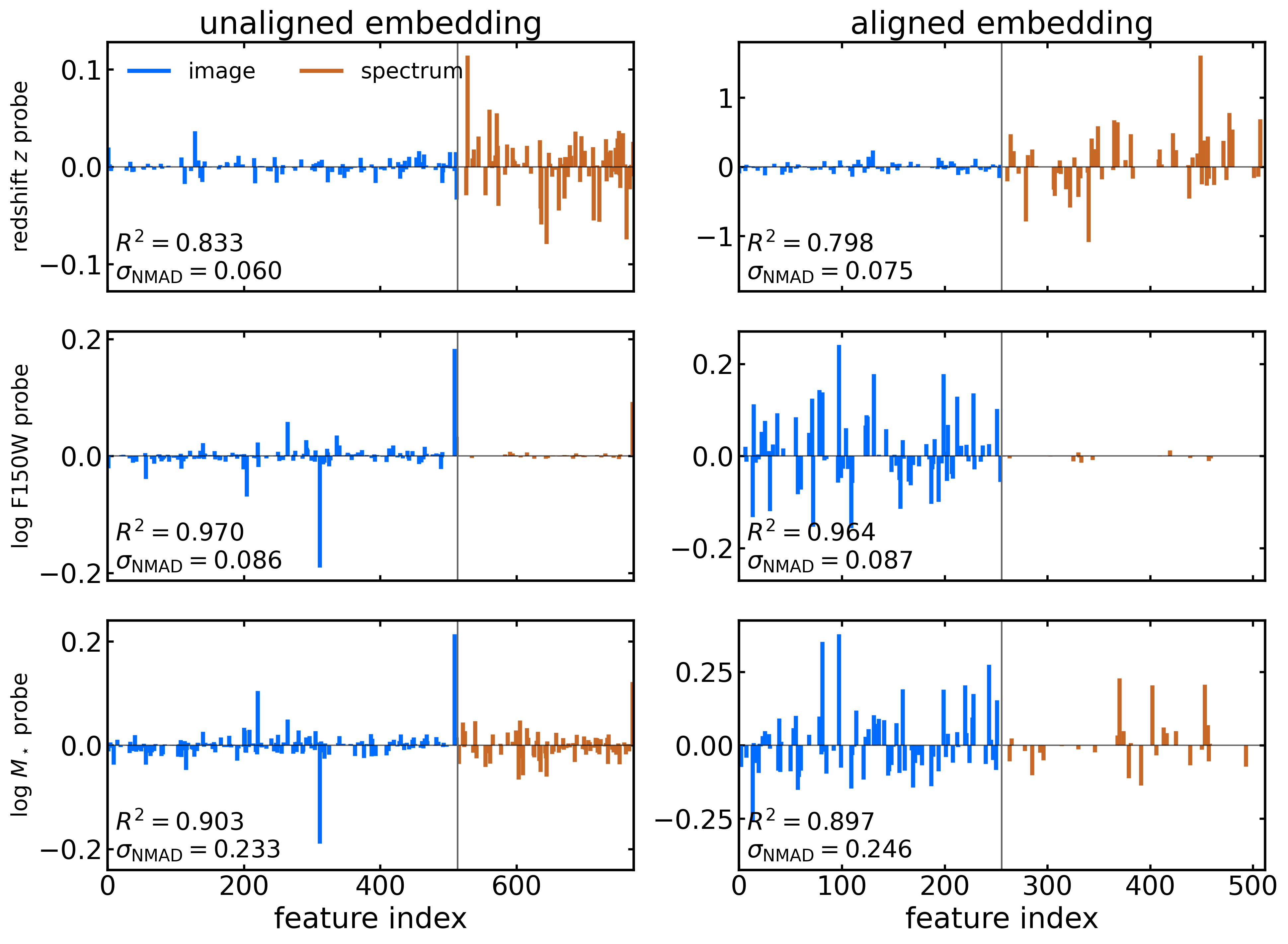}
    \caption{Weights of an $L_1$ probe on the joint image-spectrum embedding vector, before (left) and after (right) CLIP alignment, for the three targets: redshift, F150W photometry, and stellar mass (rows). Each vertical line is one non-zero coefficient, colored by the modality it belongs to; the gray line separates the image block from the spectrum block. $R^2$ and $\sigma_{\mathrm{NMAD}}$ are quoted in each panel for the split shown; repeating the fit over $40$ random $50\%/50\%$ splits at the same penalty leaves split-to-split standard deviations of at most $0.007$ and $0.006$~dex. The redshift is carried mostly by the spectrum block and the F150W flux almost entirely by the image block, whereas the stellar mass draws substantially on both.}
    \label{fig:probe_weights}
\end{figure}

\textbf{Weight breakdown of each mode:} To examine what the joint image-spectrum linear probe learns, we fit additional $L_1$ (Lasso) linear probes to the same $1<z\le3$ sample from Section~\ref{sec:results}, predicting F150W photometry, redshift, and stellar mass using both aligned and unaligned embeddings, and break down their weights. We use Lasso ($L_1$) regression to obtain a sparse solution for clearer visualization. The $L_1$ probe reproduces the ridge-regression results from Section~\ref{sec:results}: $(R^2,\ \sigma_{\mathrm{NMAD}})=(0.90,\ 0.25)$ with an $8.8\%$ outlier fraction for the CLIP-aligned joint embedding. All six fits use the same probe setup, and we plot the breakdown of linear-fit weights of those probes. The results are shown in Figure~\ref{fig:probe_weights}.

The weight breakdown explains why alignment leaves the joint embedding unchanged: stellar mass is the only target that draws substantially on both blocks, and the joint vector already contains both, so CLIP alignment has nothing further to add.

\end{document}